\documentclass[%
reprint,
superscriptaddress,
nofootinbib,
amsmath,amssymb,
aps,
10pt,
prl,
]{revtex4-2}
\usepackage{multirow}
\usepackage{graphicx}
\usepackage{dcolumn}
\usepackage{booktabs}
\usepackage{makecell}
\usepackage{threeparttable}
\usepackage{bm}

\usepackage{siunitx}
\usepackage{hyperref}
\usepackage{footnote}
\usepackage{tikz}
\usepackage{tikz-feynman}
\usepackage{float}
\allowdisplaybreaks[4]

\newcommand{\ket}[1]{| #1 \rangle}

\newcommand{\lra}[1]{\left(#1\right)}

\newcommand{\B}[0]{}
\def\xy/{$\psi(4230)$}
\def\xyy/{$\psi(4360)$}
\def\ppsi/{$\psi(4415)$}
\def\doned/{$D_1\bar{D}$}
\def\donedstar/{$D_1\bar{D}^*$}
\def\dtwodstar/{$D_2^*\bar{D}^*$} 
\def\dtwod/{$D_2^*\bar{D}$}

\usepackage{relsize}
\def\babar{\mbox{\slshape B\kern-0.1em{\smaller A}\kern-0.1em
    B\kern-0.1em{\smaller A\kern-0.2em R}}}

\usetikzlibrary{decorations.markings}
\usetikzlibrary{snakes}
\tikzset{
 photon/.style={decorate, decoration={snake}, draw=black},
    electron/.style={draw=black, postaction={decorate},
        decoration={markings,mark=at position .55 with {\arrow[draw=black]{>}}}},
    gluon/.style={decorate, draw=magenta,
        decoration={coil,amplitude=3pt, segment length=4pt}},
    scalar/.style={dashed,line width=.6pt, postaction={decorate}}
}

\newcommand{\itp}{\affiliation{CAS Key Laboratory of Theoretical Physics, Institute of Theoretical Physics,\\
Chinese Academy of Sciences, Beijing 100190, China}}
\newcommand{\ucas}{\affiliation{School of Physical Sciences, University of Chinese Academy of Sciences, Beijing 100049, China}}
\newcommand{\csu}{\affiliation{School of Physics and Electronics, Central South University, Changsha 410083, China}}

\begin{document}

\title{Prediction of a narrow exotic hadronic state with quantum numbers ${J^{PC}=0^{--}}$ }
\author{Teng Ji}\email{jiteng@itp.ac.cn}
\itp \ucas

\author{Xiang-Kun Dong}\email{dongxiangkun@itp.ac.cn}
\itp \ucas

\author{Feng-Kun Guo}\email{fkguo@itp.ac.cn}\itp \ucas

\author{Bing-Song Zou}\email{zoubs@itp.ac.cn}\itp \ucas \csu

\begin{abstract}
Lots of charmonium-like structures have been observed
{\B in the last two decades. Most of them have quantum numbers that can be formed by a pair of charm and anticharm quarks, thus it is difficult to unambiguously identify the exotic ones among them.}
{\B In this Letter, by exploiting heavy quark spin symmetry, we present a robust prediction of the hadronic molecular scenario,  where the $\psi(4230),\psi(4360)$ and $\psi(4415)$ are identified as $D\bar D_1,D^*\bar D_1$ and $D^*\bar D^*_2$ bound states, respectively. We show that a flavor-neutral charmonium-like exotic state with quantum numbers $J^{PC}=0^{--}$, denoted as $\psi_0(4360)$, should exist as a $D^*\bar D_1$ bound state.} 
{\B The mass and width of the $\psi_0(4360)$} are predicted to be $(4366\pm18)$~MeV and less than 10~MeV, respectively.
The $\psi_0(4360)$ is significant in two folds: no $0^{--}$ hadron has been observed so far, and a study of this state will enlighten the understanding of the mysterious vector mesons between 4.2 and 4.5~GeV, {\B as well as the nature of previously observed exotic $Z_c$ and $P_c$ states}. We propose that such an exotic state can be searched for in $e^+e^-\to \eta \psi_0(4360)$ and uniquely identified by measuring the angular distribution of the outgoing $\eta$ meson.

\end{abstract}

\maketitle

\emph{Introduction.---}The study on exotic states beyond the conventional quark model~\cite{GellMann:1964nj,Zweig:1964jf}, where mesons and baryons are composed of a pair of quark-antiquark($q\bar q$) and  three quarks ($qqq$), has been a focus of hadron physics in the last two decades. Quantum chromodynamics (QCD), the underlying theory that guides the formation of hadrons from quarks and gluons, does not forbid the existence of exotic configurations, 
such as multiquark states (with more than 3 valence (anti)quarks), hybrid states (with gluonic excitations in addition to the valence quarks), glueballs (only gluons) and so on. 
Many candidates of such exotic states have been observed in experiments, see Refs.~\cite{Chen:2016qju,Hosaka:2016pey,Lebed:2016hpi,Esposito:2016noz,Guo:2017jvc,Ali:2017jda,Olsen:2017bmm,Kou:2018nap,Liu:2019zoy,Brambilla:2019esw,Guo:2019twa,Yang:2020atz,Chen:2022asf,Meng:2022ozq} for recent reviews on the experimental and theoretical status of exotic states.
However, fundamental questions still remain unanswered, such as whether there is a dominant configuration for the excited hadrons and what that configuration (if any) should be.

Among the exotic states, those with exotic $J^{PC}$ are of special interests since they cannot be ordinary $q\bar q$ mesons, such as $0^{--},0^{+-},1^{-+}$ and so on. Although dozens of exotic candidates have been observed in experiments, only a few of them have exotic $J^{PC}$, including $\pi_1(1400),\pi_1(1600)$~\cite{Meyer:2015eta} and the most recently observed $\eta_1(1855)$~\cite{BESIII:2022riz,Ablikim:2022ugk}, all of which lie in the light quark sector and have $J^{PC}=1^{-+}$. Up to now, no signal of $0^{--}$ states occurs although many theoretical investigations predict the existence of such states as compact tetraquark states~\cite{Cotanch:2006wv,General:2007bk,Maiani:2014aja,Cleven:2015era,Wang:2021lkg}, hybrid states~\cite{Ishida:1991mx,General:2006ed,Govaerts:1984bk,Chetyrkin:2000tj,Huang:2016rro,Liu:2005rc}, glueballs~\cite{Qiao:2014vva,Pimikov:2016pag,Pimikov:2017bkk,Pimikov:2017xap} or a $D^*\bar D^*_0$ hadronic molecule~\cite{Shen:2010ky}. One should notice that the above predictions may have large uncertainties and some of them are still controversial, even problematic. For example, the QCD sum rules concluded that no $0^{--}$ tetraquark state exists below 2~GeV~\cite{LEE:2020eif,Jiao:2009ra}; the $D_0^*$ is too wide to form a bound state~\cite{Filin:2010se,Guo:2011dd} and its mass listed in the Review of Particle Physics (RPP)~\cite{Workman:2022ynf} is too high (see Ref.~\cite{Du:2020pui} and references therein).

Hadronic molecules, {\B as analogues of light nuclei, are composite systems of a few hadrons. Being close to the thresholds of their components, they can be studied using nonrelativistic effective field theory. They can be distinguished from other exotic configurations by investigating long-distance processes involving the constituents~\cite{Guo:2017jvc} and exploiting approximate symmetries of QCD such as heavy quark spin symmetry (HQSS)~\cite{Cleven:2015era}.} Most of the experimental candidates of exotic states with small widths are found to be possible hadronic molecules, such as the $\chi_{c1}(3872)$~\cite{Choi:2003ue}, the $Z_c(3900)^\pm$~\cite{Ablikim:2013mio,Liu:2013dau,Ablikim:2013xfr}, the $P_c$ states~\cite{Aaij:2019vzc} and the $T_{cc}^+$~\cite{LHCb:2021auc,LHCb:2021vvq}, see Refs.~\cite{Guo:2017jvc,Dong:2021juy,Dong:2020hxe,Meng:2022ozq} for reviews and general discussions. Besides, the $\psi(4230)$, $\psi(4360)$ and $\psi(4415)$ are good candidates of hadronic molecules of $1^{--}$ $D\bar D_1$, $D^*\bar D_1$ and $D^*\bar D^*_2$, respectively~\cite{Wang:2013cya,Wang:2013kra,Ma:2014zva,Cleven:2015era,Hanhart:2019isz,Anwar:2021dmg} (throughout this Letter, the $D_1$ refers to the narrow $D_1(2420)$ listed in the RPP~\cite{Workman:2022ynf}, and $D\bar D_1$ means a linear combination of $D\bar D_1$ and their antiparticles with certain $C$-parity and similar for others), especially after noticing the remarkable feature that 
\begin{align}
    m_{\psi(4360)}-m_{\psi(4230)}&\approx m_{D^*}-m_{D},\\
    m_{\psi(4415)}-m_{\psi(4360)}&\approx m_{D_2^*}-m_{D_1},
\end{align}
which is a natural consequence of HQSS where the low energy interaction between hadrons is independent of the spins of heavy quarks. 
In fact, the interactions in these channels are the most attractive ones among all the narrow charm-(anti)charm meson pairs from exchanging the light vector mesons~\cite{Dong:2021juy,Dong:2021bvy}, and thus these states could be the deepest bound hadronic molecules in the hidden-charm and double-charm meson-meson sectors.

In the heavy quark limit of $m_c\to\infty$, $D$ and $D^*$ belong to the same spin multiplet $H$ with the angular momentum of the light degrees of freedom $s_{\ell}=1/2$, and $D_1,D_2^*$ belong to the multiplet $T$ with $s_{\ell}=3/2$. Heavy quark spin partners of the $\psi(4230)$ have been estimated using a constant interaction from the vector-meson dominance model in the exploratory study of Ref.~\cite{Dong:2021juy}, among which there are four isoscalar states with exotic quantum numbers: a $0^{--}$ $D^*\bar D_1$ molecule (denoted as $\psi_0$) around 4.4~GeV, and three $1^{- +}$ $D\bar D_1$, $D^*\bar D_1$ and $D^*\bar D_2^*$ molecules (denoted as $\eta_{c1}$) from about 4.3 to 4.5~GeV. They can be searched for in hadron and $e^+e^-$ collisions. In $e^{+}e^-$ collisions below 5~GeV, within the current reach of the BESIII experiment, the $1^{- +}$ can be produced through $e^+e^-\to \gamma \eta_{c1}$, while the $\psi_0$ can be produced in reactions with hadronic final states $e^+e^-\to \eta \psi_{0}$.
Therefore, it is timely to carefully investigate the $\psi_0$, which does not mix with ordinary charmonia and provides a unique portal to understand the vector states in the energy region between 4.2 and 4.5~GeV.
In this Letter, we show that the existence of the explicitly exotic $\psi_0$ is robust in the molecular picture of the vector states $\psi(4230)$, $\psi(4360)$ and $\psi(4415)$, and it can be searched for in electron-positron collisions with an unambiguous signature.

\begin{table}[t]
	\centering
	\caption{\label{tab:molecules}Hadronic molecules considered in this work and their possible experimental candidates. The binding energies {\B $E_B\equiv m_1 + m_2 - M$, where $M$ and $m_{1,2}$ are the masses of the hadronic molecule and its constituents, respectively,} of the $1^{--}$ states are obtained by the experimental masses of their candidates and that of the $\psi_0$ is the prediction in this Letter. The values of the thresholds and $E_B$ are in units of MeV.
    }
			\begin{tabular}{l|c*{2}{c}c}
				\Xhline{1.0pt}
				Molecule & Components & $J^{PC}$ & Threshold & $E_B$ \\
				\Xhline{0.4pt}
				$\psi(4230)$&  $\frac{1}{\sqrt2}(D\bar D_1-\bar D D_1)$&$1^{--}$ & $4287$ &${\B 67\pm15}$\\
                \Xhline{0.4pt}
                $\psi(4360)$&  $\frac{1}{\sqrt2}(D^*\bar D_1-\bar D^* D_1)$&$1^{--}$ & $4429$ &${\B 62\pm14}$\\
                \Xhline{0.4pt}
                $\psi(4415)$&  $\frac{1}{\sqrt2}(D^*\bar D_2-\bar D^* D_2)$&$1^{--}$ & $4472$ &${\B 49\pm4}$\\
                \Xhline{0.4pt}
                {$\psi_0$}&  { $\frac{1}{\sqrt2}(D^*\bar D_1+\bar D^* D_1)$}&{ $0^{--}$} & {4429} &{$\bm{{\B 63\pm 18}}$}\\
				\Xhline{1.0pt}
			\end{tabular}
\end{table}

\emph{Framework.---}The flavor wave functions of the $\psi(4230)$, $\psi(4360)$ and $\psi(4415)$ as $1^{--}$ molecules, and $\psi_0$ as a $0^{--}$ molecule are listed in Table~\ref{tab:molecules}, where we have adopted the following charge conjugation conventions,
\begin{align}
\mathcal C\ket{D}&=\ket{\bar D},\quad \mathcal C \ket{D^*}=-\ket{\bar D^*},\notag\\
\mathcal C\ket{D_1}&=\ket{\bar D_1},\quad \mathcal C \ket{D_2^*}=-\ket{\bar D_2^*}\label{eq:C}.
\end{align}

In the near-threshold energy region, the interactions between charmed mesons can be described with a nonrelativistic effective field theory, and at leading order there are four independent constant contact terms for the $S$-wave interactions between the $H$ and $T$ multiplets (for each possible isospin)~\cite{Guo:2017jvc}. 
In the lack of data to fix these contact terms, their values may be estimated with the resonance saturation model by considering the exchange of light mesons~\cite{Ecker:1988te,Epelbaum:2001fm}.

In the following we first focus on the $t$-channel exchanges and then discuss the $u$-channel pion-exchange corrections (the contribution of the $u$-channel exchanges of other mesons are much weaker than the $t$-channel ones~\cite{Dong:2019ofp,Dong:2021juy}). 
We consider the exchange of light vector ($V$) and pseudoscalar ($P$) mesons by keeping the momentum dependence in the Yukawa potentials, which can be regarded as a way of resumming part of higher order contributions in the momentum expansion. 

{\B The three meson-meson $1^{--}$ channels listed in the second column of Table~\ref{tab:molecules} can couple with one another, and the scattering amplitude by the $t$-channel $V$ and $P$ exchanges can be expressed as
\begin{align}
   \B \mathcal M_{ij}={\frac{A^V_{ij}}{\bm q^2+m_V^2}+\frac{A^P_{ij}}{\bm q^2+m_P^2}+{c_V} B^V_{ij}}+{c_P} B^P_{ij},\label{eq:MVP}
\end{align}
where $\bm q$ is the transferred 3-momentum and $i,j=1,2,3$ denote channels. They are derived using Lagrangians satisfying HQSS, SU(3) flavor symmetry and chiral symmetry (and hidden local symmetry for light vectors), constructed in Refs.~\cite{Wise:1992hn,Falk:1992cx,Grinstein:1992qt,Casalbuoni:1992gi,Casalbuoni:1992dx,Casalbuoni:1996pg} and collected in Ref.~\cite{Dong:2021juy}. The coefficients $A^{V,P}_{ij},B^{V,P}_{ij}$ can be expressed in terms of coupling constants with phenomenologically known values~\cite{Bando:1987br,Isola:2003fh,Chen:2019asm,Wang:2021yld}, see the Supplemental Material. The first two terms of the amplitude correspond to Yukawa potentials contributing to the long and mid-range interaction, while the last two are short-range constant contact terms.
It turns out that the $t$- and $u$-channel exchanges of  $V$ and $P$ mesons produce four different contact terms; the number matches that of the leading order terms from HQSS analysis~\cite{Guo:2017jvc} mentioned above.
Since the contact terms produce ultraviolet (UV) divergence in the Lippmann-Schwinger equation (LSE), they receive scale dependence from renormalization. 
Therefore, we have introduced two scale-dependent factors $c_V(\Lambda)$ and $c_P(\Lambda)$ to the constant terms in Eq.~\eqref{eq:MVP} serving as counterterms (the constant terms from the $u$-channel exchanges produce another two).

The nonrelativistic potential in momentum space reads
\begin{align}
     V_{ij}=-\frac1{\Pi_{\alpha=1}^4\sqrt{2m_\alpha}}\mathcal M_{ij},
\end{align}
with $m_\alpha$ the mass of the initial or final {\B particles}. 
The potential for the $0^{--}$ system is similar with the $1^{--}$ channels with different coefficients $A^{V,P}_{0},B^{V,P}_{0}$ (also shown in the Supplemental Material) and the same parameters $c_V$ and $c_P$ due to HQSS.}

Bound states are obtained by solving the LSE,
\begin{align}
T_{i j}(E ; \boldsymbol{k}^{\prime}&, \boldsymbol{k})=V_{i j}\left(\boldsymbol{k}^{\prime}, \boldsymbol{k}\right)\notag\\
&+\sum_{n} \int \frac{{\rm d}^{3} \boldsymbol{l}}{(2 \pi)^{3}} \frac{V_{i n}\left(\boldsymbol{k}^{\prime}, \boldsymbol{l}\right) T_{n j}(E ; \boldsymbol{l}, \boldsymbol{k})}{E-\boldsymbol{l}^{2} /\left(2 \mu_{n}\right)-\Delta_{n 1}+i \epsilon},\label{eq:LSE}
\end{align}
where $\bm k$ and $\bm{k'}$ are the 3-momenta of the initial and final states in the center-of-mass (c.m.) frame, $\mu_n$ is the reduced mass of the $n$-th channel, $E$ is the energy relative to threshold of the first channel and $\Delta_{n1}$ is the difference between the $n$-th threshold and the first one. A Gaussian form factor is introduced to regularize the UV divergence,
\begin{align}
    V_{i j}\left(\boldsymbol{k}^{\prime}, \boldsymbol{k}\right)\to V_{i j}\left(\boldsymbol{k}^{\prime}, \boldsymbol{k}\right)e^{-\bm{q}^2/\Lambda^2},
\end{align}
with $\bm{q}=\bm{k}-\bm{k}'$ and $\Lambda$ the cutoff.

The parameters $c_V(\Lambda)$ and $c_P(\Lambda)$ at a given $\Lambda$ are determined by reproducing the binding energies of the three $1^{--}$ molecular candidates, {\B i.e., by minimizing the $\chi^2$ function defined as
\begin{align}
        \chi^2=\sum_{i}\lra{\frac{E_{B,ii}-E_{{\rm exp},ii}^{\rm cen}}{E_{{\rm exp},ii}^{\rm err}}}^2,\label{eq:chi2}
\end{align}
where $E_{B,ii}$ is the calculated binding energy of the $i$-th channel bound state depending on $c_V(\Lambda)$ and $c_P(\Lambda)$, and $E_{{\rm exp},ii}^{\rm cen,\, err}$ is the corresponding experimental central value and error, as shown in the last column in Table~\ref{tab:molecules}.}

\begin{figure}[t]
    \centering
    \includegraphics[width=\linewidth]{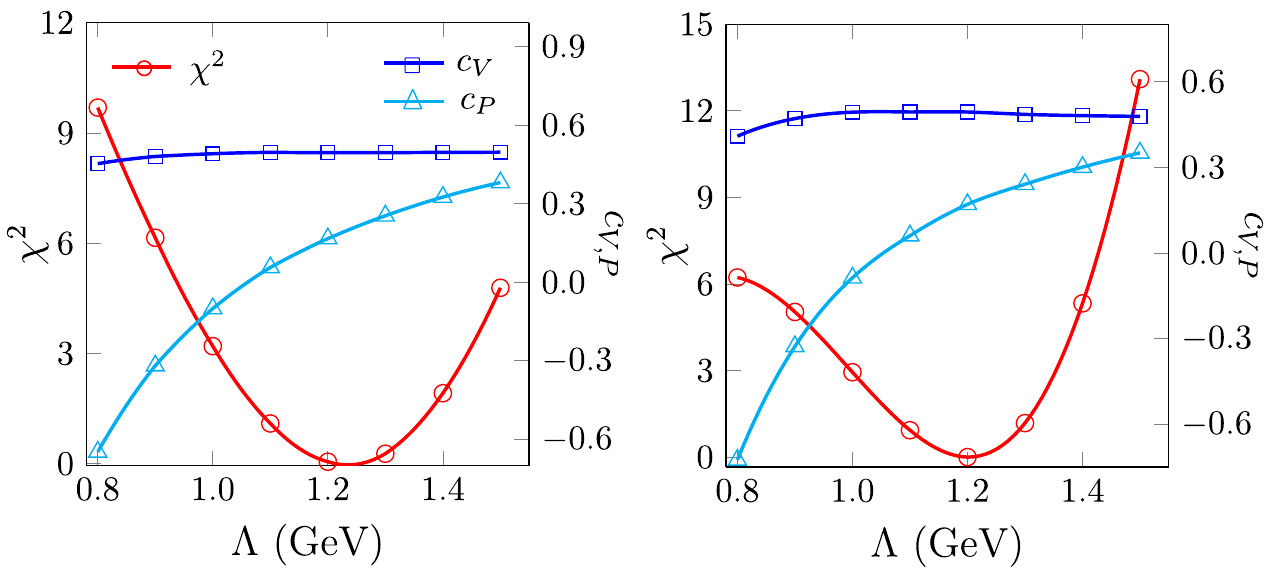}
    \caption{The minimized $\chi^2$ as a function of $\Lambda$ and the corresponding $c_{V,P}(\Lambda)$ for the single-channel ({left}) and coupled-channel ({right}) cases. 
    }
    \label{fig:chi2cd}
\end{figure}

\emph{$t$-channel results.---}Let us first focus on the single-channel case by turning off the off-diagonal elements of the potential matrix $V_{ij}$. By minimizing the $\chi^2$ in Eq.~(\ref{eq:chi2}) for a given $\Lambda$, which is chosen in the phenomenologically reasonable range of $0.8\sim 1.5$~GeV, we obtain the results shown in the left plot of Fig.~\ref{fig:chi2cd}. It is clear that when $\Lambda\approx 1.2$~GeV, we can find suitable $c_V=0.50,c_P=0.18$ reproducing  the experimental central values and the corresponding binding energy of $\psi_0$ is $(72.4 \pm 17.4)$~MeV, where the error is estimated by setting $\chi^2=1$ for $\Lambda=1.2$~GeV.

\emph{Coupled-channel effects.---}After turning on the off-diagonal $V_{ij}$, the poles corresponding to the $\psi(4360)$ and $\psi(4415)$ will move to the complex plane on the unphysical Riemann sheets (RSs) due to the opening of the lower \doned/ channel. 
However, it turns out that the coupled-channel effects are negligible and the pole locations are very close to the real axis (the imaginary part at $\Lambda=1.2$~GeV is less than $1$~MeV). We take the real parts of the complex pole locations as the corresponding binding energies {\B (up to a sign)} and the results are shown in the right plot of Fig.~\ref{fig:chi2cd}. The best solution is still located at $\Lambda\approx1.2$~GeV and the predicted binding energy of $\psi_0(4360)$ is ${72.4}$~MeV. The difference from the single-channel result, $\sim0.1$ MeV, is much less than the estimated uncertainty from the experimental errors, see the Supplemental Material for more comparisons. 
Therefore, we conclude that the coupled-channel effects are negligible.

\emph{$u$-channel pion exchange and 3-body effects.---} Although the contribution from the $u$-channel exchange is usually small, the  $u$-channel exchanged pion can go on-shell {\B in the current case}, which means that it contributes to the longest-range interaction and the intermediate 3-body channel will introduce additional cuts to the scattering amplitude and result in nonzero decay widths of the predicted molecules. Thus, we take the $D^*\bar D_1$ single channel as an example to carefully investigate such 3-body effects to the pole positions.

{\B It is known that for the $D_1(2420)$, although dominated by an $s_\ell=3/2$ state, the $S$-wave contribution to the decay width of $D_1\to D^*\pi$ is sizeable~\cite{Falk:1992cx,Guo:2020oqk}. The $S$-wave and $D$-wave coupling constants for $D_1D^*\pi$ are determined to be $|g_S|= 2.0\ \rm GeV^{-1}$ and $|g_D|= 4.9\ {\rm GeV}^{-2}$ from the decay widths of $D_1$ and $D_2^*$. The $D$-wave $D_1D^*\pi$ coupling would lead to new UV divergence in the $u$-channel pion exchange and calls for more counterterms. To avoid this issue, we consider only the $S$-wave coupling with two different values of $g_S$: $g_{S0}=2.0$~GeV$^{-1}$ as given above and $g_{S1}=\sqrt{31/12}\, g_{S0}$ to mimic the total width of $D_1$. As the $D$-wave vertex is of higher order in the momentum expansion than the $S$-wave one, the real $u$-channel pion-exchange contribution should live between these two extreme cases.}


\begin{figure}
    \centering
    \includegraphics[width=0.75\linewidth]{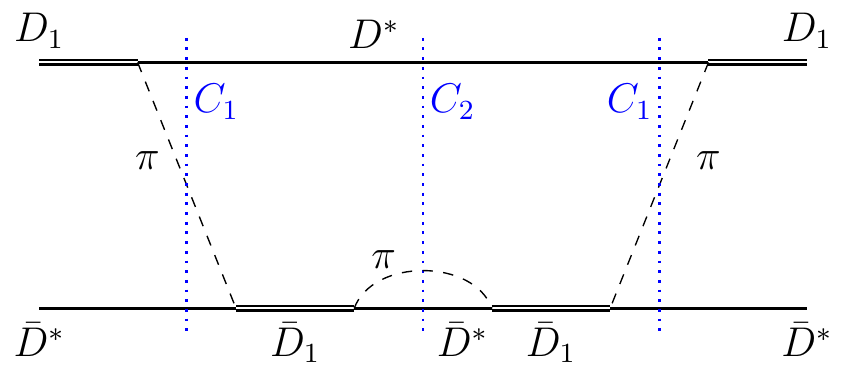}
    \caption{{\B An illustration of the three-body cuts (vertical dotted lines) introduced by the simultaneous onshellness of the intermediate particles, with $C_1$ for the $u$-channel pion exchange and $C_2$ for the $D_1$ self-energy (decay width).}}
    \label{fig:3bcuts}
\end{figure}

{\B The $D^*\bar D^*\pi$ 3-body channel enters the problem in two aspects~\cite{Doring:2009yv,Baru:2011rs,Du:2021zzh}, as illustrated by the two kinds of cuts ($C_1$ and $C_2$) in Fig.~\ref{fig:3bcuts},}
which should be properly treated when searching for poles on the unphysical RS. The details can be found in the Supplemental Material and here we only show the final results, as listed in Table~\ref{tab:pole}. We find that the $D_1$ self-energy gives the molecule a width smaller than that of $D_1$ and has little influence on the binding energy, and the $u$-channel pion exchange has influence on both the real and imaginary parts of the pole position: the imaginary parts from the above two contributions interfere constructively for the $\psi(4360)$ and destructively for the $\psi_0$; the binding energies change by $\lesssim 10\ \rm MeV$, within the errors of the $t$-channel results. 
As discussed above, the real 3-body effects should live between those of $g_S=g_{S0}$ and $g_{S1}$
since the $D$-wave coupling in Eq.~(\ref{eq:LSD}) is of higher order in the derivative expansion than the $S$-wave one.
Thus, we conclude that the $\psi_0$ has a mass of $(4366\pm18)$~MeV, where the central value is obtained by averaging the results of $\psi_0$ with $g_S=g_{S0}$ and $g_{S1}$ and the uncertainty sums in quadrature half their difference and the one from the $t$-channel fitting (that in the second row of Table~\ref{tab:pole}).

The existence of lower channels which are not considered here can increase the widths, which are twice the absolute values of the imaginary parts of the poles listed in Table~\ref{tab:pole}. In particular, the $S$-wave $J/\psi(\psi')\pi\pi$ and $P$-wave $D^{(*)}\bar D^{(*)}$ should be crucial to bring the width { of $\psi(4360)$} to  $(96\pm7)$~MeV~\cite{Workman:2022ynf} measured by experiments~\cite{BaBar:2012hpr,Belle:2014wyt,BESIII:2016bnd,BESIII:2017tqk}. 
On the contrary, the $\psi_0$ cannot decay into $J/\psi(\psi')\pi^+\pi^-$, $D\bar D$ or $D^*\bar D^*$, and its width should be significantly smaller than that of the $\psi(4360)$. An estimate of the decay width by considering $\psi_0\to D^*\bar D_1\to D\bar D^*$ through the $V$ and $P$ exchanges leads to about $\lesssim1$~MeV, {\B and the partial width of the 3-body decay mode $\psi_0\to D^*\bar D^*\pi$ as given in Table~\ref{tab:pole} lies in the range between 0.6 and 2.2~MeV}. Consequently, we expect the total width of the $\psi_0$ to be well below $10$~MeV.

\begin{table*}[t]
\caption{Pole positions relative to the $D^*\bar D_1$ threshold in units of $\rm MeV$ with $c_V=0.50,c_P=0.18$ from the single $t$-channel fitting. {\B The real part corresponds to the mass relative to the  $D^*\bar D_1$ threshold, and the absolute value of the imaginary part corresponds to half the width.} The uncertainties of $t$-channel {\B results are from minimizing the $\chi^2$ function in Eq.~\eqref{eq:chi2}}. {\B ``$C_2$"} means the $D_1$ self-energy considered while the $u$-channel pion exchange not and {\B ``$C_1\& C_2$"} means both contributions included.}\label{tab:pole}
\begin{tabular}{c|cc|cc}
\Xhline{0.8pt}
System                & \multicolumn{2}{c|}{$1^{--}$}                        & \multicolumn{2}{c}{$0^{--}$}                        \\ \hline 
$t$-channel           & \multicolumn{2}{c|}{$-63.5\pm 13.8$}                         & \multicolumn{2}{c}{$-72.4\pm 17.4$}                         \\ \hline
$g_S$ & \multicolumn{1}{c|}{$g_{S0}$}           & $g_{S1}$ & \multicolumn{1}{c|}{$g_{S0}$}          & $g_{S1}$ \\ \hline
{\B $C_2$}   & \multicolumn{1}{c|}{$-61.5-3.5i$}  & $-61.5-9.2i$    & \multicolumn{1}{c|}{$-70.0-3.5i$} & $-70.0-8.9i$    \\ \hline
{\B $C_1\& C_2$  }         & \multicolumn{1}{c|}{$-65.8-6.6i$} & $-73.1-14.2i$    & \multicolumn{1}{c|}{$-65.8-0.30i$} & $-59.4-1.1i$   \\ \Xhline{0.8pt}
\end{tabular}
\end{table*}

\emph{Experimental search.---}The $\psi_0$, {\B  with exotic quantum numbers $0^{--}$, cannot couple to $c\bar c$. Thus, its production rate in $B$ decays through the weak process $b\to c\bar c s$ should be tiny, contrary to the particles like the $\chi_{c1}(3872)$ which can be produced through  $c\bar c$ via the $\bar s\gamma_\mu(1-\gamma_5)b\, \bar c \gamma^\mu(1-\gamma_5)c$ operators~\cite{Braaten:2004ai,Meng:2005er}. However, it} can be searched for in electron-positron collisions in final states such as $D\bar D^*$ and  $J/\psi (\psi') \eta$ {\B via the processes $e^+e^-\to \eta^{(\prime)} \psi_0(4360)$ with the $\eta^{(\prime)}$ and $\psi_0(4360)$ in a $P$-wave.
Although the production of $\eta\psi_0(4360)$, whose threshold is about 4.9~GeV, at the current BESIII experiment should be highly suppressed due to the limited phase space and $P$-wave suppression, it is promising at the  upcoming  BEPCII-U upgrade~\cite{BESIII:2022mxl}, which has an energy reach up to 5.6~GeV and has a higher luminosity than the current BEPCII, and at Belle-II~\cite{Belle-II:2018jsg}.}
Given that the $e^+e^-\to \pi^{+} D^{0} D^{*-}$ cross section is as large as about 0.4~nb at 4.6~GeV, and the integrated luminosity of BESIII at 4.95~GeV is 0.16~fb$^{-1}$~\cite{BESIII:2022ulv}, there is a high chance for the $\psi_0(4360)$ to be found in the $D\bar D^{*}$ final state of $e^+e^-\to \eta D \bar D^{*}$ {\B at higher energies}.

However, this process may always be accompanied by $e^+e^-\to\eta\psi(4360)$. The decay channels of $\psi_0$, such as $D\bar D^*$ and $J/\psi\eta$, are also shared by the $\psi(4360)$. Furthermore, the masses of these two resonances are similar. Therefore, we need to identify an observable that is unique in distinguishing the $\psi_0$ from the $\psi(4360)$, and the  distribution of the angle between the outgoing $\eta$ and the $e^+e^-$ beam in the laboratory frame, denoted as $\theta$, fulfills the requirement.

\begin{figure}[t]
    \centering
    \includegraphics[width=0.9\linewidth]{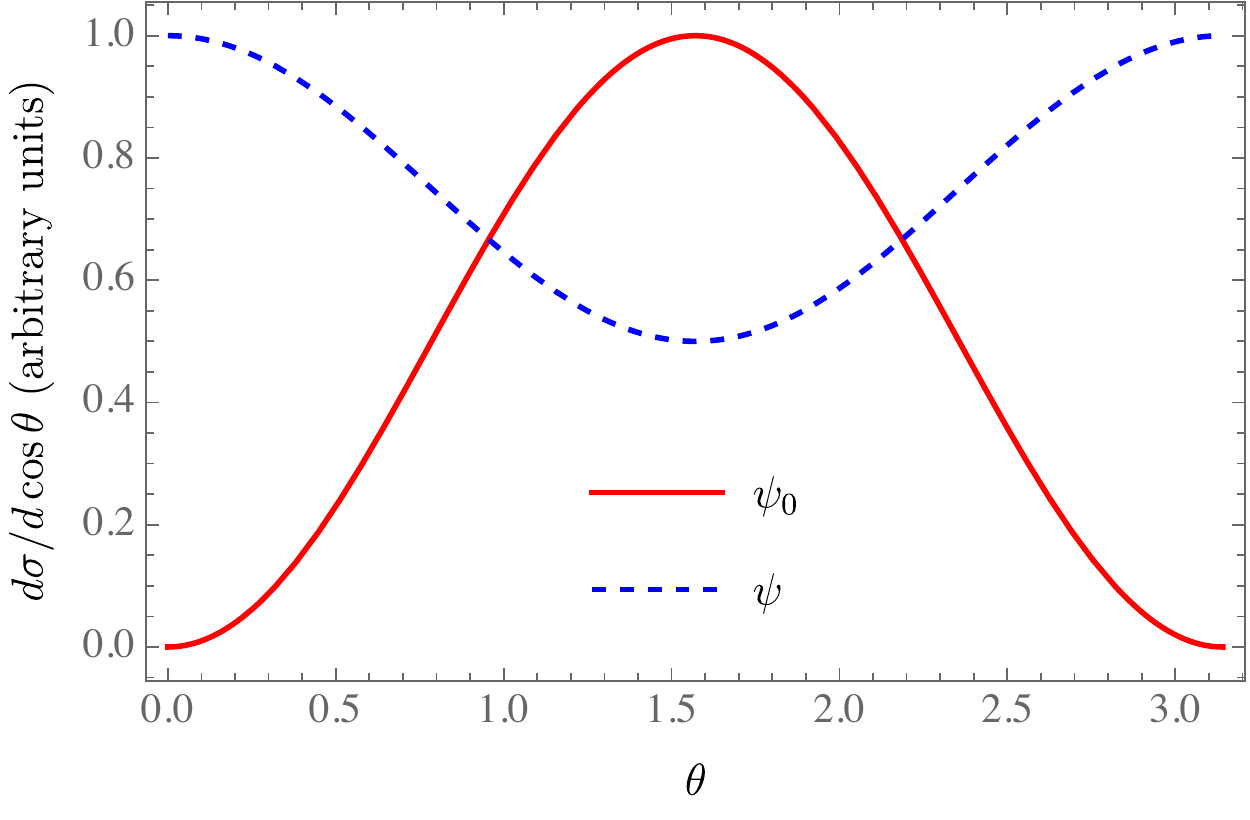}
    \caption{Angular distributions of $e^+e^-\to\eta\psi_{0}$ and $\eta\psi$. $\theta$ is the angle between the outgoing $\eta$ and initial $e^+e^-$ beam. The distributions are in arbitrary units and the maxima are normalized to 1.}
    \label{fig:ang-dis}
\end{figure}

For $e^+e^-\to \gamma^* \to \eta(p_1)\psi_0(p_2)$ and $\eta(p_1)\psi(p_2)$, the amplitudes $\mathcal M_0$ and $\mathcal M_1$ have the following forms,
    \begin{align}
        \mathcal{M}_0&\propto\epsilon(\gamma^*)\cdot q ,\\
        \mathcal{M}_1&\propto\epsilon_{\alpha\beta\gamma\delta}\epsilon^\alpha(\gamma^*)\epsilon^{*\beta}(\psi)P^\gamma q^\delta ,
    \end{align}
where $P=p_1+p_2$ and $q=p_1-p_2$. Because the intermediate virtual photon $\gamma^*$ from $e^+e^-$ annihilations at high energy is transversely polarized, summing over the initial and final polarizations leads to totally different angular distributions for $e^+e^-\to\eta\psi$ and $e^+e^-\to\eta\psi_0$, as shown in Fig.~\ref{fig:ang-dis}, and the $\psi_0$ signal can be clearly distinguished from that of the $\psi(4360)$. 

\emph{Conclusion and outlook.---}The existence of a $0^{--}$ $D^*\bar D_1$ bound state $\psi_0$ is a natural consequence in the molecular scenario of the $\psi(4230), \psi(4360)$ and $\psi(4415)$.  
Being explicitly exotic, it does not mix with charmonium states.
We have shown that the existence of the $\psi_0$ is robust against coupled-channel and 3-body pion-exchange effects.
The mass and width of the $\psi_0$ are predicted to be $(4366\pm18)$~MeV and $\lesssim10$~MeV, respectively. We may denote such a state as $\psi_0(4360)$.

It is promising to search for the $\psi_0(4360)$ in $e^+e^-$ collisions through the process $e^+e^-\to \eta \psi_0(4360)$. The angular distribution provides an unambiguous signature to distinguish the explicitly exotic $\psi_0(4360)$ from states of other possible quantum numbers, such as a vector state in the same mass range. 
Moreover, the width of the $\psi_0(4360)$ is expected to be much smaller than that of the $\psi(4360)$.

{\B }

So far no $0^{--}$ meson has been observed.
Being a robust prediction of the hadronic molecular model, the $\psi_0(4360)$ will provide a unique opportunity to infer 
the internal structure of the vector mesons in the mass range between 4.2 and 4.5~GeV, which has been a riddle since the discovery of the $\psi(4260)$~\cite{BaBar:2005hhc}. {\B Its possible observation
would also play a crucial role to establish the hadonic molecular nature for $Z_c$, $P_c$ and many other exotic hadronic states.} 

\medskip

We are grateful to the fruitful discussions with Vadim Baru, Meng-Lin Du, Jia-Jun Wu, Mao-Jun Yan and Chang-Zheng Yuan. This project is supported in part by the Chinese Academy of Sciences under Grants No. XDB34030000 and No. XDPB15; by the National Natural Science Foundation of China (NSFC) under Grants No. 12125507, No. 11835015, No. 12047503, and No. 11961141012; and by the NSFC and the Deutsche Forschungsgemeinschaft (DFG) through the funds provided to the Sino-German Collaborative Research Center TRR110 “Symmetries and the Emergence of Structure in QCD” (NSFC Grant No. 12070131001, DFG Project-ID 196253076).

\bibliography{refs}
\appendix
\section{Supplemental Material}
\subsection{$t$-channel potentials}
The amplitudes for $D^{(*)}\bar D_{1,2}^{(*)}$ scattering via the $V$ and $P$ exchanges can be expressed in Eq.~(\ref{eq:MVP}) and the coefficients take the following expressions,
\begin{align}
    A_{11}^V&=-4\beta\beta_2g_V^2m_Dm_{D_1}-\frac{20}{9}\beta\lambda_2g_V^2m_Dm_\rho^2, \\
    B_{11}^V&=\frac{20}{9}\beta\lambda_2g_V^2m_D, \\
    A_{11}^P&=B_{11}^P=0, \\
    A_{22}^V&=-4\beta\beta_2g_V^2m_Dm_{D^*}-\frac49g_V^2m_\rho^2(6\beta_2\lambda m_{D_1}\notag\\
    &\ \ \ \ +5\beta\lambda_2m_{D^*}+20\lambda\lambda_2m_{D_1}m_{D^*}), \\
    B_{22}^V&=\frac49g_V^2(6\beta_2\lambda m_{D_1}+5\beta\lambda_2m_{D^*}+20\lambda\lambda_2m_{D_1}m_{D^*}),\\
    A_{22}^P&=-\frac32\frac{10gk}{9f_\pi^2}m_{D_1}m_{D^*}m_\pi^2=-\frac34m_\pi^2B_{22}^P,\\
    A_{33}^V&=-4\beta\beta_2g_V^2m_{D^*}m_{D_2}-\frac83g_V^2m_\rho^2(\beta\lambda_2m_{D^*}\notag\\
    &\ \ \ \ +\beta_2\lambda m_{D_2}-6\lambda\lambda_2m_{D^*}m_{D_2}),\\
    B_{33}^V&=\frac83g_V^2(\beta\lambda_2m_{D^*}+\beta_2\lambda m_{D_2}-6\lambda\lambda_2m_{D^*}m_{D_2}),\\
    A_{33}^P&=-\frac32\frac{2gk}{f_\pi^2}m_{D_2}m_{D^*}m_\pi^2=-\frac34m_\pi^2B_{33}^P,\\
    A_{12}^V&=\frac{20\sqrt2}{9}g_V^2\lambda\lambda_2m_{D_1}(m_{D^*}+m_D)m_\rho^2\notag\\
    &=-m_\rho^2 B_{12}^V,\\
    A_{12}^P&=-\frac{3}{2}\frac{10\sqrt2gk}{9f_\pi^2}m_{D_1}\sqrt{m_{D^*}m_D}m_\pi^2\notag\\
    &=-\frac{3}{4}m_\pi^2B_{12}^P,\\
    A_{13}^V&=\frac{8\sqrt{10}}{3}g_V^2\lambda\lambda_2m_{D_1}m_Dm_\rho^2=-m_\rho^2 B_{13}^V,\\
    A_{13}^P&=-\frac{3}{2}\frac{2\sqrt{10}gk}{9f_\pi^2}\sqrt{m_Dm_{D^*}m_{D_1}m_{D_2}}m_\pi^2\notag\\
    &=-\frac{3}{4}m_\pi^2B_{32}^P,\\
    A_{23}^V&=-\frac{16\sqrt{5}}{9}g_V^2\lambda\lambda_2\sqrt{m_{D_1}m_{D_2}}m_{D^*}m_\rho^2=-m_\rho^2 B_{23}^V,\\
    A_{23}^P&=\frac{3}{2}\frac{2\sqrt{5}gk}{9f_\pi^2}\sqrt{m_{D_1}m_{D_2}}m_{D^*}m_\pi^2=-\frac{3}{4}m_\pi^2B_{32}^P,\\
    A_0^V&=-4\beta\beta_2g_V^2m_Dm_{D^*}-\frac49g_V^2m_\rho^2(6\beta_2\lambda m_{D_1}\notag\\
    &\ \ \ \ +5\beta\lambda_2m_{D^*}+40\lambda\lambda_2m_{D_1}m_{D^*}),\\
    B_0^V&=\frac49g_V^2(6\beta_2\lambda m_{D_1}+5\beta\lambda_2m_{D^*}+40\lambda\lambda_2m_{D_1}m_{D^*}),\\
    A_0^P&=A_{22}^P,\quad    B_0^P=B_{22}^P.
\end{align}
Note that the isospin factors for isospin-0 have been taken into account.
For the $P$ exchange, the Yukawa term contains only the pion-exchange contributions while the corresponding counterterm contains contributions from both the pion and the $\eta^{(\prime)}$ mesons. 
The numerical values of coupling constants are $g=-k=0.59$~\cite{Isola:2003fh,Falk:1992cx}, $\beta=-\beta_2=0.9$~\cite{Isola:2003fh}, $g_V=5.8$~\cite{Bando:1987br}, $\lambda=-\lambda_2=0.56$~\cite{Chen:2019asm,Wang:2021yld} and $f_\pi=132$~MeV is the pion decay constant. The masses of the involved particles are taken from the RPP~\cite{Workman:2022ynf}.

\subsection{Comparisons between the single- and coupled-channel results}

\begin{figure}[tb]
    \centering
    \includegraphics[width=0.8\linewidth]{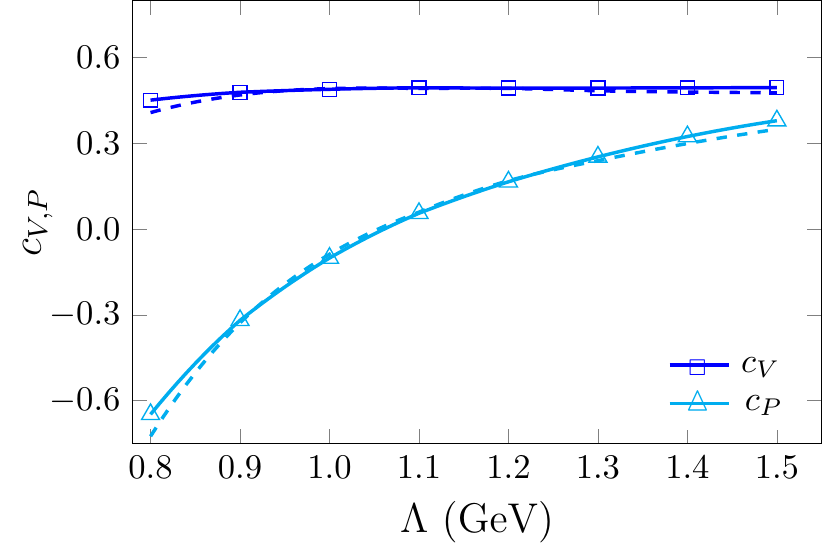}
    \caption{Comparison between $c_{V,P}$ of single-channel (solid) and coupled-channel (dashed) results with only $t$-channel meson-exchange interaction.}
    \label{fig:cdcom}
\end{figure}

\begin{figure*}[htb]
        \centering
         \includegraphics[width=0.485\linewidth]{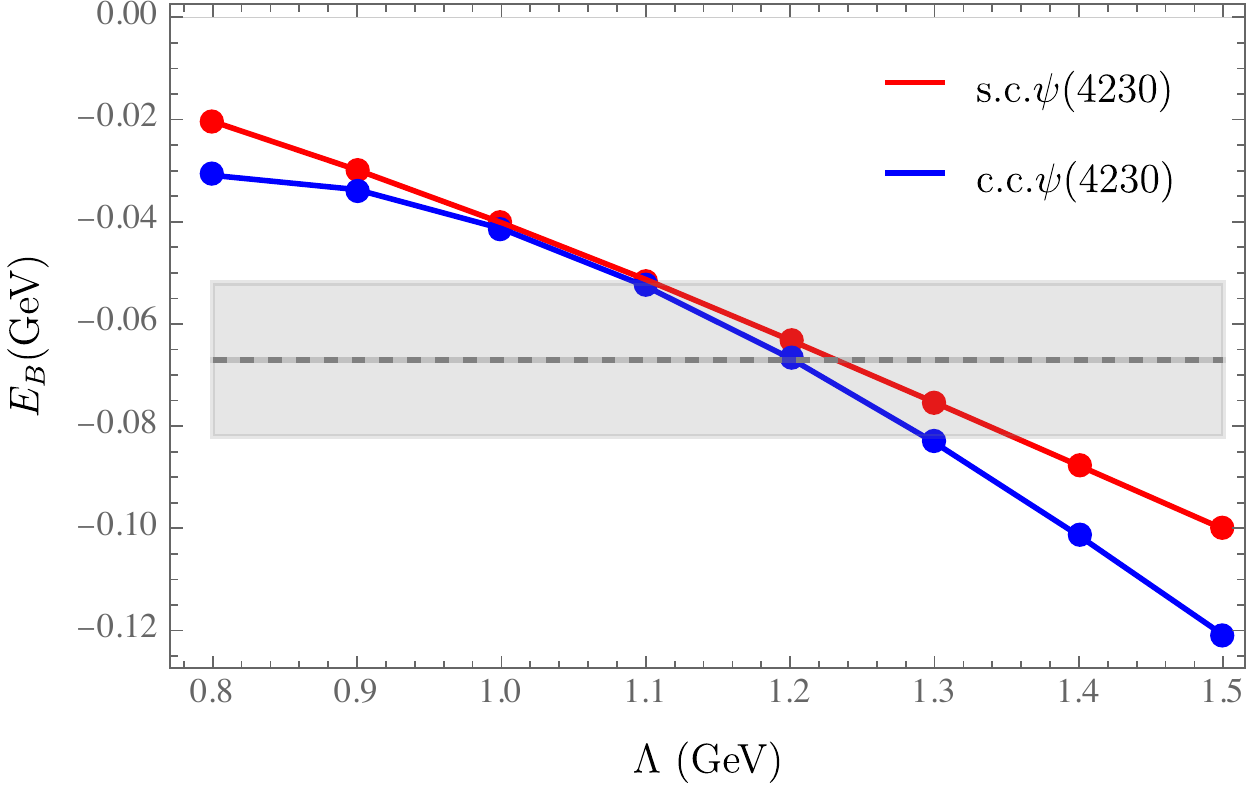}\hfill
        \includegraphics[width=0.485\linewidth]{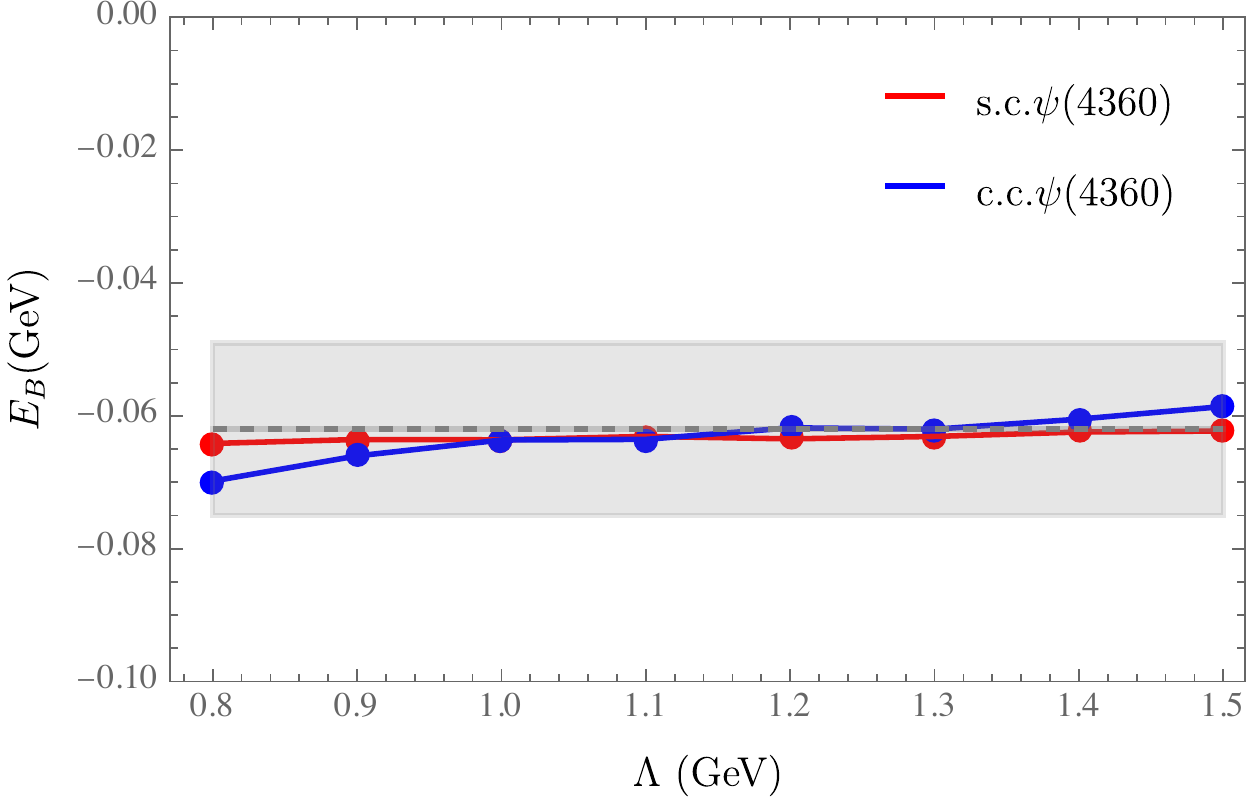}
        \includegraphics[width=0.485\linewidth]{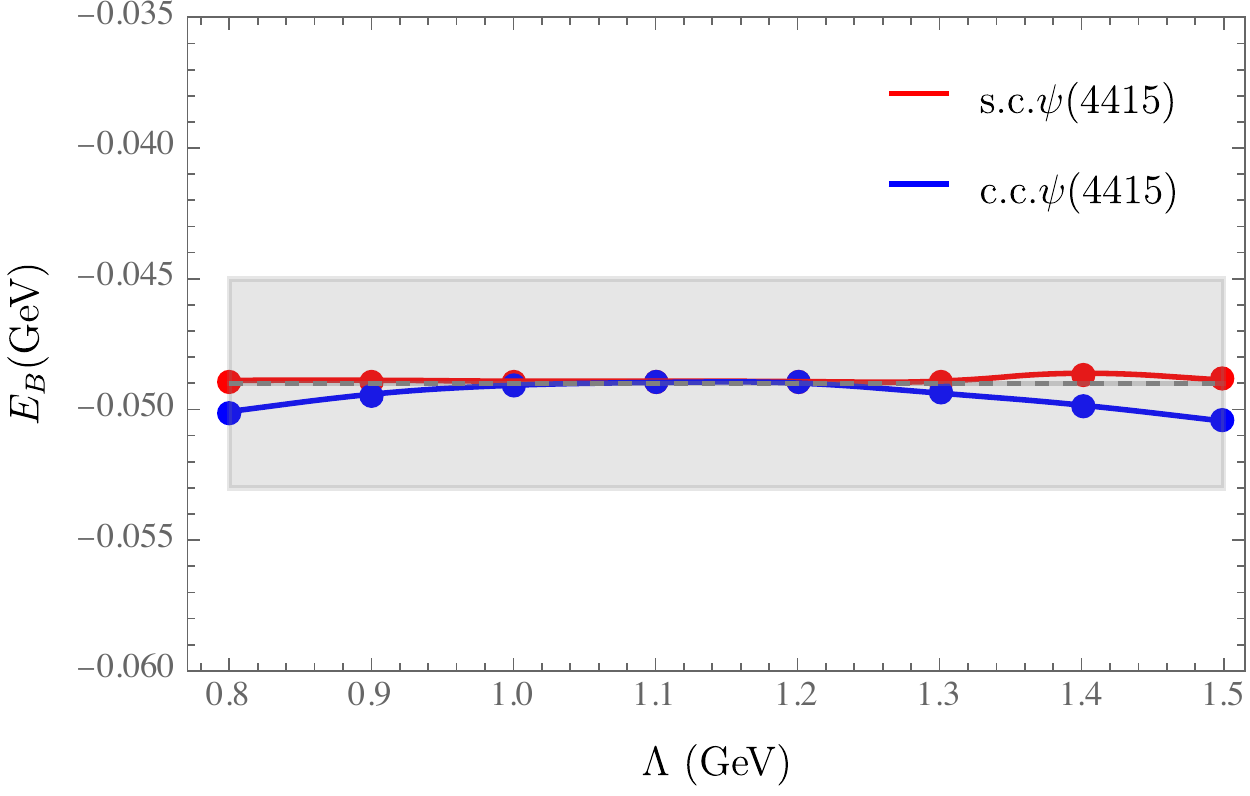}\hfill
        \includegraphics[width=0.485\linewidth]{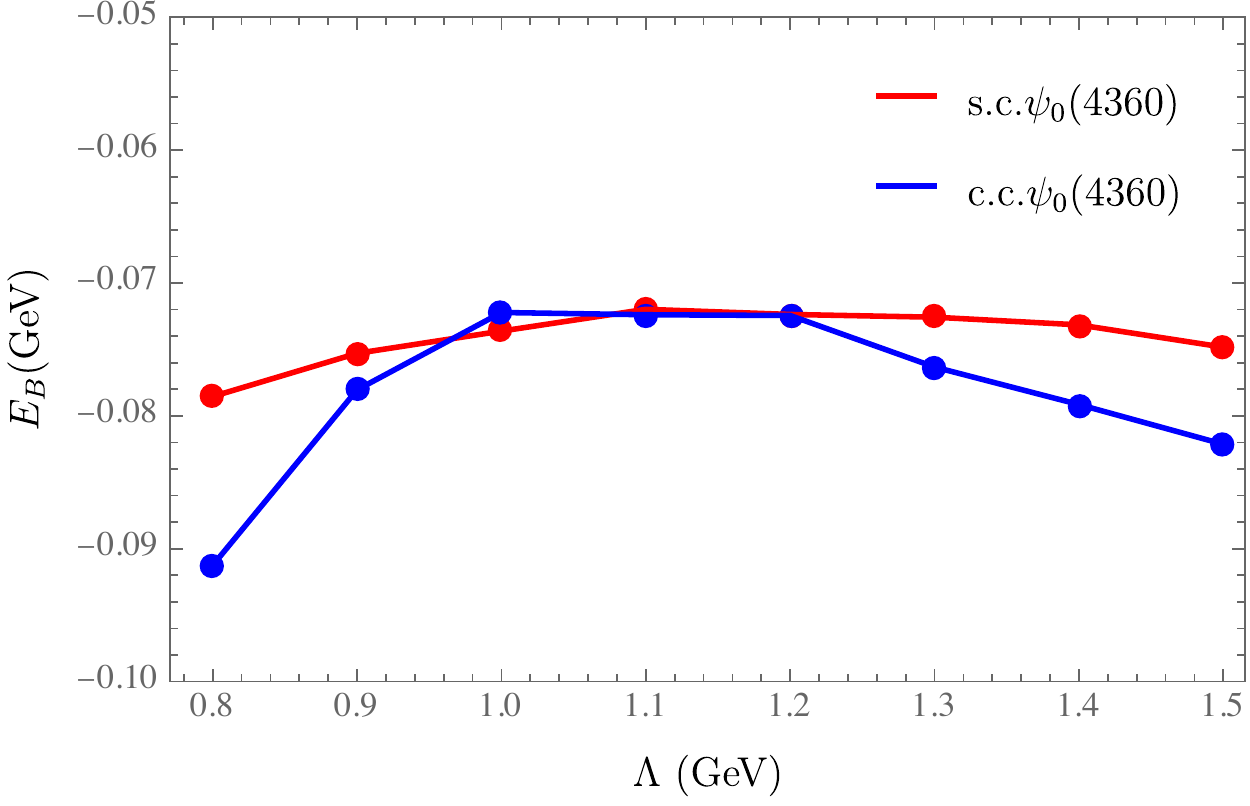}
        \caption{Comparison of the single-channel (s.c.) and coupled-channel (c.c.) results for binding energies of the $\psi(4230)$, $\psi(4360)$, $\psi(4415)$ and $\psi_0(4360)$ with only the $t$-channel meson-exchange interaction. The dashed lines and shadows represent the experimentally measured central values and errors.}
        \label{fig:ebs}
\end{figure*}

Here we compare the values of $c_{V,P}$ explicitly in Fig.~\ref{fig:cdcom}. We can see that for $\Lambda=1.0\sim1.3$~GeV, there are almost no differences between single-channel and coupled-channel results. The binding energies corresponding to these parameters are shown in Fig.~\ref{fig:ebs} and the predicted binding energy of $\psi_0(4360)$ are robust against $\Lambda$.

\subsection{Analytical properties of the amplitudes including the 3-body channel}
{\B 
 The $D_1D^*\pi$ coupling can be derived from the Lagrangian
\begin{align}
    \mathcal{L}_{D_1D^*P}=&\,g_S \sqrt{m_{D_1}m_{D^*}}D_1^{\mu}\partial_\nu PD^{*\dagger}_\mu v^\nu\notag\\
    &+g_D\sqrt{m_{D_1}m_{D^*}}\left[3D_{1}^{\mu}\partial_{\mu}\partial_{\nu}PD^{*\dagger\nu}\right.\notag\\
    &\left.-D_{1}^{\alpha}\left(\partial^2P-\partial_{\mu}\partial_{\nu}Pv^\mu v^\nu\right)D^{*\dagger}_\alpha\right]+\rm h.c.,\label{eq:LSD}
\end{align}
with $g_S= 2.0\ \rm GeV^{-1}$ and $g_D= 4.9\ {\rm GeV}^{-2}$, where the latter is fixed from $D_2^*$ width and the former from reproducing the $D_1$ width together with the $g_D$ term. Correspondingly,
the $S$- and $D$-wave decay widths of $D_1(2420)$ are around $12$ and $19$~MeV, respectively. 

For the $S$-wave coupling in Eq.~(\ref{eq:LSD}), the transition amplitude of $D^*\bar D_1\to D_1\bar D^*$ reads
\begin{align}
    \mathcal M_u=\frac{g_S^2}{4}(m_{D_1}^2-m_{D^*}^2)^2\frac{1}{q^2-m_\pi^2+i\epsilon} . \label{eq:Mu}
\end{align}
As discussed in the main text, we consider only the $S$-wave coupling with two different values of $g_S$: $g_{S0}=2.0$~GeV$^{-1}$ as given above and $g_{S1}=\sqrt{31/12}\, g_{S0}$ to mimic the total width of $D_1$. 

{\B The $D^*\bar D^*\pi$ 3-body channel enters the problem in two aspects~\cite{Baru:2011rs,Du:2021zzh}.} First, the propagator of the exchanged pion in LSE reads
\begin{align}
   \frac1{q^2-m_\pi^2+i\epsilon}\to\frac1{2E(m_\pi,\bm q)}\left(\frac{1}{d_1}+\frac1{d_2}\right)\label{eq:propu}
\end{align}
where 
\begin{align}
    d_i=\sqrt{s}-E(m_\pi,\bm q)-E(m_i,\bm k)-E(m_i,\bm{k'})
    \label{eq:di}
\end{align}
with $m_1=m_{D^*},m_2=m_{D_1},\sqrt s=E+m_{D^*}+m_{D_1}$ and $E(m,\bm p)=\sqrt{m^2+\bm p^2}$. Second, the $D^*\pi$ loop contributes to the $D_1$ self-energy, leading to an energy-dependent width of $D_1$, 
\begin{align}
            \Gamma_{D_1}(E,\bm{l} )=\frac{g_S^2}{4}(m_{D_1}^2-m_{D^*}^2)^2\frac{p_{\rm cm}}{8\pi m_{D^*\pi}^2}, \label{eq:Gam}
\end{align}
where $m_{D^*\pi}$ is the invariant mass of $D^*\pi$ and $p_{\rm cm}$ is the magnitude of the 3-momentum of $D^*/\pi$ in their c.m. frame, determined by
\begin{align}
           \sqrt{ m_{D^*\pi}^2+\bm l^2}+\sqrt{ m_{D^*}^2+\bm l^2}&=E+m_{D^*}+m_{D_1},\\
           \sqrt{ m_{D^*}^2+p_{\rm cm}^2}+\sqrt{ m_{\pi}^2+p_{\rm cm}^2}&=m_{D^*\pi}.
\end{align}
The $i\epsilon$ term in Eq.~(\ref{eq:LSE}) should be replaced with $i\Gamma_{D_1}(E, \bm l)/2$.
}

The on-shellness of the propagating intermediate particles will introduce cuts to the scattering amplitudes, which are analytical functions of the complex energy except for these cuts and possible poles. Usually, the cut introduced by the intermediate component particles, say $D^*\bar D_1$ in the $\psi(4360)$ case, is chosen from the threshold to positive infinity, called the right-hand cut (RHC) while those from the exchanged particles are  called the left-hand cuts (LHCs). Here we give detailed discussions on the cuts in the $D^*\bar D_1$ scattering amplitude including the $D^*\bar D^*\pi$ 3-body channel.

\begin{figure}[t]
        \centering
        \includegraphics[width=0.9\linewidth]{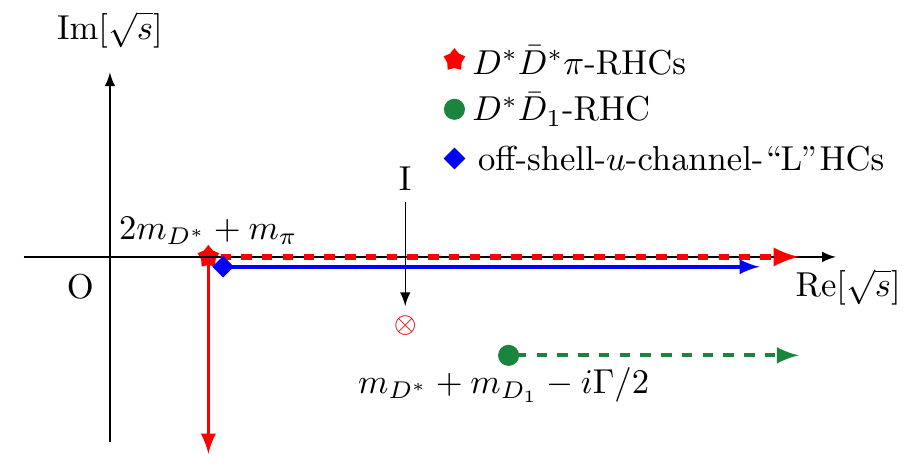}
        \caption{The cuts encountered in the $D^*\bar D_1$ system. The off-shell $u$-channel ``L"HC is located on the real axis and moved away slightly for better illustration. For $D^*\bar D^*\pi$ RHCs and off-shell $u$-channel ``L"HCs, only that with $|\bm l|=0$ or $|\bm k|,|\bm {k'}|=0$ is shown. {\B The pole positions of the $\psi$ and $\psi_0$ are marked by the red $\otimes$, which can be reached from the physical RS (denoted by I) by crossing the ``L"HCs from the off-shell $u$ channel pion exchange.}}
        \label{fig:Allcuts}
    \end{figure}

The RHC in the $D^*\bar D^*\pi$ (with $\bar D^*\pi$ from $\bar D_1$) channel originates from the $p_{\rm cm}$ in $\Gamma_{D_1}(E,\bm l)$ (see Eq.(\ref{eq:Gam})) and starts from $s=s_0(|\bm l
|)$ to positive infinity by choosing a cut of the square root function in $p_{\rm cm}$ so that 
\begin{equation}
    \begin{array}{ll}
       {\rm Im} [p_{\rm cm}]\ge0  & \text{on 1st RS},\\
        {\rm Im} [p_{\rm cm}]<0 & \text{on 2nd RS}.
    \end{array}\label{eq:RSdef}
\end{equation}
Note that the branch point $s_0(|\bm l|)$ is a moving point from ${s_0(0)}=(2m_{D^*}+m_\pi)^2$ to the right as $|\bm l|$ increases. See the red-dashed line in Fig.~\ref{fig:Allcuts} for the cut with $|\bm l|=0$.
Due to the finite decay width of $D_1$, one of the branch points of the $D^*\bar D_1$ RHC is located at $\sqrt{s}=m_{D^*}+m_{D_1}-\frac{i}{2}\Gamma_{D_1}$ on the 2nd RS, instead of at the nominal threshold on the real axis. See the green line in Fig.~\ref{fig:Allcuts}.

To see the LHCs, we need to analyze the poles of the meson-exchange potentials (both on-shell and off-shell for the initial and final scattered particles) before the partial wave projection. For the on-shell $t$-channel one, the cut lies at $s\in (-\infty,s_1]$ with $s_1<(m_{D^*}+m_{D_1})^2$. While for the off-shell one with $|\boldsymbol{k}|,|\boldsymbol{k'}|\ge0$, there is no cut. The branch points for the $u$-channel pion exchange are determined by $d_1=0$ at $z\equiv\cos\widehat{\bm k\cdot \bm{k}}{}\bm{'}=\pm1$ with $d_1$ given in Eq.~\eqref{eq:di}, which translates to
\begin{align}
            \sqrt{m_{D_1}^2+\bm k^2} =&\, \sqrt{m_{D^*}^2+\bm k^{\bm{\prime}2}}\notag\\
            &+\sqrt{m_{\pi}^2+\bm k^2+\bm k^{\bm{\prime}2}\pm2|\bm k||\bm{k'}|}.
\end{align}
For the on-shell case where $|\bm k|=|\bm {k'}|$, 
it is actually a RHC  (still called ``L"HC in the following) from a point $s_2$ above threshold to positive infinity. For the off-shell case with $|\bm k|,|\bm {k'}|\ge0$, one of the branch points, $s_3(|\bm k|,|\bm{k'}|)$, is moving from $s_3(0,0)=(2m_{D^*}+m_\pi)^2$ to the right as $|\bm k|$ and $|\bm{k'}|$ increase and the other is positive infinity, again a RHC (``L"HC). See the blue line in Fig.~\ref{fig:Allcuts} for the cut with $|\bm k|,|\bm{k'}|=0$.

\begin{figure}[t!]
        \centering
        \includegraphics[width=\linewidth]{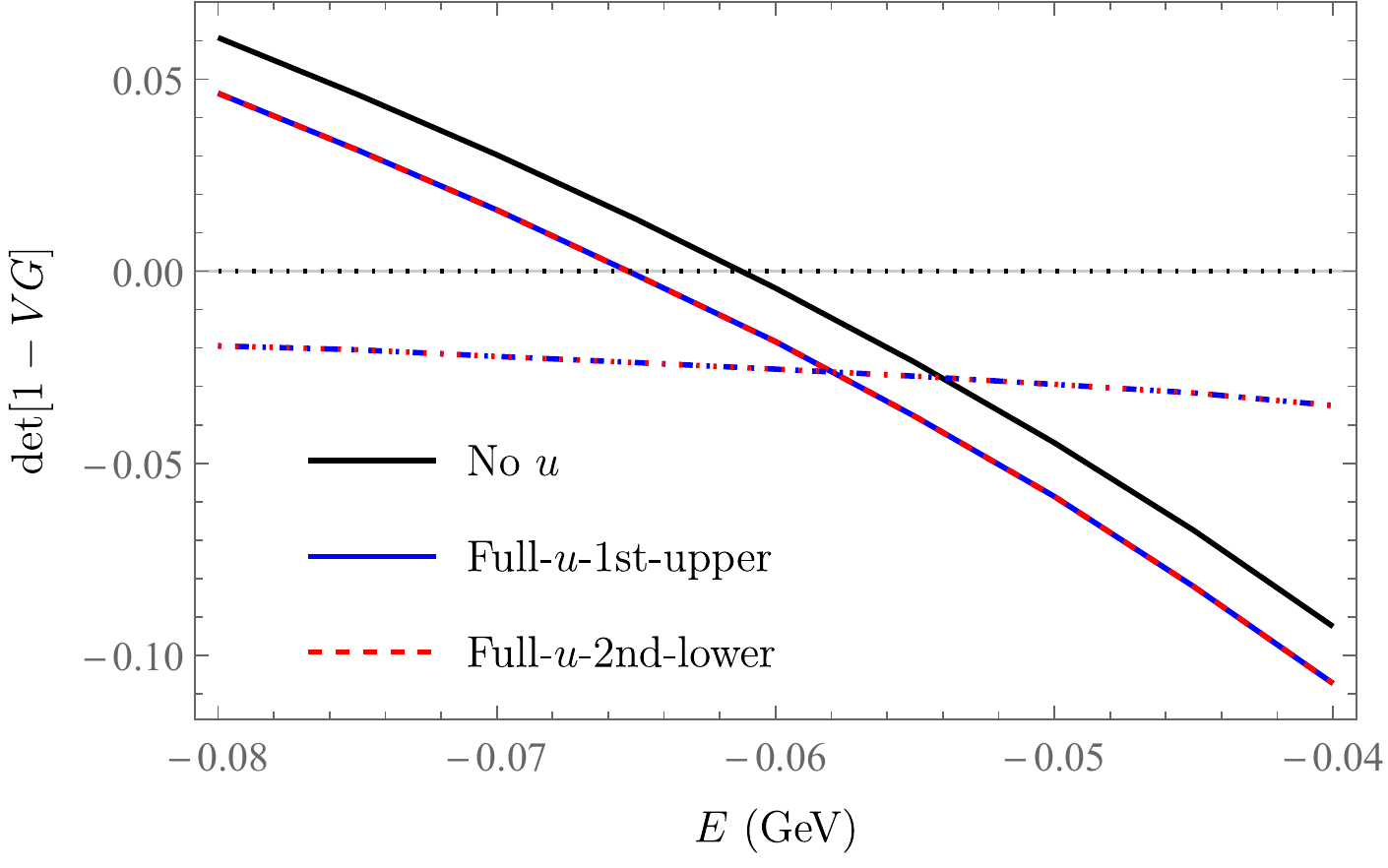}
        \caption{An illustration of the connection between the 1st-upper ($E+i10^{-7}$~GeV) and 2nd-lower ($E-i10^{-7}$~GeV) RSs where the solid and dashed (dotted and dotdashed) lines are the real (imaginary) parts. ``No $u$" means only $t$-channel interaction considered and in turn no cuts while ``Full $u$" means that the $u$-channel interaction and $D_1$ self-energy are both included. The blue curves are covered by the red ones and hence they are connected correctly. }
        \label{fig:cross}
\end{figure}
To search for poles corresponding to the $\psi_0(4360)$ and $\psi(4360)$, which are now located on the unphysical RS marked by the red $\otimes$ in Fig.~\ref{fig:Allcuts} and close to the physical axis on the first RS, one has to perform the analytical continuation properly. 
\begin{itemize}
    \item $t$- and $u$-channel cuts when the scattered particles are on-shell: They will not contribute by construction when solving LSE and therefore, they are not shown in Fig.~\ref{fig:Allcuts}.
    \item The off-shell $u$-channel ``L"HCs with $|\bm k|,|\bm {k'}|\ge0$: When performing the $S$-wave projection, i.e. the integral over $z$ from $-1$ to $1$ along the real axis, the pole of the integrand, $z_0(E,|\bm k|,|\bm{k'}|)$, will move from the lower half $z$-plane to the upper one when $E$ moves from physical axis to the unphysical region where the pole of $T$ matrix is located. If $z_0(E,|\bm k|,|\bm{k'}|)$ cross over the integral path, the result will have discontinuity. To avoid crossing over these cuts, we deform the integral path of $z$ from a straight line $-1\to1$ to a polyline $-1\to -1+i a\to 1+ia\to 1$ with $a$ a sufficiently large positive number.
    \item The RHC of $D^*\bar D_1$: This one is on the unphysical sheet, beyond the region of the possible pole and hence of no interest here.
    \item The $D^*\bar D^*\pi$ RHCs: If we use the cuts defined in Eq.~(\ref{eq:RSdef}), it is found that the values of $|1-VG|$ (an intermediate function in solving LSE, whole roots are the pole positions of the $T$ matrix) in the 1st-upper and 2nd-lower RSs do not match. The reason is that when integrating $|\bm l|$ from $0$ to $+\infty$ along the real axis, the argument of the square root function in $p_{\rm cm}$ would go across its branch point from right to left, see the Fig. 7 in Ref~\cite{Doring:2009yv}. Therefore we have to change the integral path to avoid this discontinuation. In our calculation, we have chosen a proper cut of the square root which lies parallel to the negative imaginary axis to realize a smooth continuation~\cite{Doring:2009yv,Du:2021zzh}. See the red-solid line in Fig.~\ref{fig:Allcuts} for the modified cut with $|\bm l|=0$. 
\end{itemize}

After the above treatments, the values of $|1-VG|$ are continued smoothly from the physical axis to the 2nd RS, see Fig.~\ref{fig:cross} for an illustration.
\end{document}